\documentclass[journal=jctc,manuscript=article,layout=twocolumn]{achemso}

\usepackage{graphicx}
\usepackage{amssymb}
\usepackage{amsmath}
\usepackage{color}

\newcommand{\R}{\mathbf{r}}
\newcommand{\fv}{\mathbf{f}}
\newcommand{\funcder}[2]{\frac{\delta #1}{\delta #2}}
\newcommand{\deri}[2]{\frac{\partial #1}{\partial #2}}

\author{Eduardo Fabiano}
\email{eduardo.fabiano@cnr.it}
\affiliation{Institute for Microelectronics and Microsystems (CNR-IMM), Via Monteroni, Campus Unisalento, 73100 Lecce, Italy}
\affiliation{Center for Biomolecular Nanotechnologies @UNILE, Istituto Italiano di Tecnologia, Via Barsanti, I-73010
Arnesano, Italy}
\author{Szymon \'Smiga}
\affiliation{Institute of Physics, Faculty of Physics, Astronomy and Informatics,
Nicolaus Copernicus University, Grudziadzka 5, 87-100 Torun, Poland}
\author{Sara Giarrusso}
\affiliation{Department of Theoretical Chemistry and Amsterdam Center for Multiscale Modeling, Faculty of Science, Vrije Universiteit, De Boelelaan 1083, 1081HV Amsterdam, The Netherlands}
\author{Timothy J. Daas}
\affiliation{Department of Theoretical Chemistry and Amsterdam Center for Multiscale Modeling, Faculty of Science, Vrije Universiteit, De Boelelaan 1083, 1081HV Amsterdam, The Netherlands}
\author{Fabio Della Sala}
\affiliation{Institute for Microelectronics and Microsystems (CNR-IMM), Via Monteroni, Campus Unisalento, 73100 Lecce, Italy}
\affiliation{Center for Biomolecular Nanotechnologies @UNILE, Istituto Italiano di Tecnologia, Via Barsanti, I-73010
Arnesano, Italy}
\author{Ireneusz Grabowski}
\affiliation{Institute of Physics, Faculty of Physics, Astronomy and Informatics,
Nicolaus Copernicus University, Grudziadzka 5, 87-100 Torun, Poland}
\author{Paola Gori-Giorgi}
\affiliation{Department of Theoretical Chemistry and Amsterdam Center for Multiscale Modeling, Faculty of Science, Vrije Universiteit, De Boelelaan 1083, 1081HV Amsterdam, The Netherlands}
\date{\today}
 
\title{Investigation of the exchange-correlation potential of functionals based on the adiabatic connection interpolation}

\begin{document}

\begin{abstract}
We have studied the correlation potentials produced by various adiabatic connection models (ACM)
for several atoms  and molecules. The results have been compared to accurate reference potentials (coupled cluster and 
quantum Monte Carlo results) as well as to state-of-the-art ab initio DFT approaches.
We have found that all the ACMs yield correlation potentials that exhibit a correct behavior,
quite resembling scaled second-order G\"orling-Levy (GL2) potentials, and including most of the physically meaningful features 
of the accurate reference data.
The behavior and contribution of the strong-interaction limit potentials has  also been 
investigated and discussed. 
\end{abstract}

%\maketitle

\section{Introduction}
The study of the exchange-correlation (XC) functional and the development of efficient and accurate approx- imations to it are among the main research topics in density functional theory (DFT). Within this theoreti- cal framework, the XC functional describes in fact all the quantum effects of the electron-electron interaction and it finally determines the accuracy of the overall com- putational procedure.

Over the years many different XC approximations have been developed
\cite{functionals_chap,dellasala16} and they are conventionally 
organized on the so called Jacob's ladder of DFT \cite{dftladder}.
To the highest rung of the ladder belong functionals that depend
on the Kohn-Sham orbitals and eigenvalues. Some
examples of these are the random-phase approximation \cite{jiang07,rpa}, 
double-hybrids functionals~\cite{Gri-JCP-06,DHOEPSmiga2016}
and the ab initio DFT methods \cite{bartlett:2005:abinit2,abinitiodft,IGccpt2}.

Another class of high-level XC approximations is
the one of functionals based on interpolating the adiabatic connection integrand between its weak and strong-interaction limits.
These functionals use as starting point the adiabatic connection formula
\cite{harris74,langreth75,gunnarsson76,savin03}
\begin{equation}
E_{xc} = \int_0^1W_\lambda[\rho]d\lambda\ ,
\end{equation}
where $\rho$ is the electron density, $\lambda$ 
is the interaction strength, and 
$W_\lambda[\rho]=\langle\Psi_\lambda[\rho]|\hat{V}_{ee}|\Psi_\lambda[\rho]\rangle - U[\rho]$
is the density-fixed linear adiabatic
connection integrand, with $\Psi_\lambda[\rho]$ being the wave
function that minimizes $\hat{T}+\lambda\hat{V}_{ee}$ 
while yielding the density $\rho$ ($\hat{T}$ and $\hat{V}_{ee}$
are the kinetic and electron-electron interaction operators, respectively), 
and $U[\rho]$ being the Hartree energy.
The task of constructing an XC approximation is then translated to
the one of developing a proper 
approximation for the density-fixed linear adiabatic
connection integrand $W_\lambda$ \cite{isi} by interpolating between its 
known exact asymptotic behaviors in the weak- and strong-interaction
limits \cite{gl2,SeiPerLev-PRA-99,seidl07,gorigiorgi09}, i.e.
\begin{eqnarray}
W_{\lambda\rightarrow0}[\rho] & \sim & W_0[\rho] + \lambda W'_0[\rho] + \cdots \label{eq:lambda0} \\ 
\label{eq:lambdainf} 
W_{\lambda\rightarrow\infty}[\rho] & \sim & W_\infty[\rho] + \frac{1}{\sqrt{\lambda}} W'_\infty[\rho] + \cdots\ ,
\end{eqnarray}
with
\begin{equation}
W_0[\rho] = E_x[\rho]\; , \; W'_0[\rho]=2\,E_c^{\rm GL2}[\rho]\ ,
\end{equation}
where $E_x$ is the exact exchange, $E_c^{\rm GL2}$ is
the second-order G\"orling-Levy (GL2)~\cite{gl2}
correlation energy, $W_\infty[\rho]$ is the indirect
part of the minimum expectation value of the
electron-electron repulsion in a given density \cite{seidl07}, 
and $W'_\infty[\rho]$ is
the potential energy of coupled zero-point
oscillations \cite{gorigiorgi09}. 

The functionals $W_\infty[\rho]$ and $W'_\infty[\rho]$ 
have a highly nonlocal density dependence, captured by the strictly-correlated electrons (SCE) limit \cite{seidl07,gorigiorgi09,MalMirCreReiGor-PRB-13}, and their exact evaluation in general cases is a non-trivial problem.  Notice that while the form of the leading term $W_\infty[\rho]$ in Eq.~(\ref{eq:lambdainf}) rests on recent mathematical proofs \cite{Lew-arxiv-17,CotFriKlu-arxiv-17}, the zero-point term $W'_\infty[\rho]$ is a very reasonable conjecture that has been confirmed numerically in simple cases \cite{GroKooGieSeiCohMorGor-JCTC-17}, but lacks a rigorous proof. 
The $\lambda\to\infty$ functionals can also be approximated by the much cheaper semilocal gradient expansions (GEA) derived within the point-charge-plus-continuum (PC) model \cite{seidl00}
\begin{eqnarray}
\label{e5}
{W}_\infty^{\rm PC}[\rho] & = & \int \left[A\rho^{4/3}(\R)+B\frac{|\nabla\rho(\R)|^2}{\rho^{4/3}(\R)}\right]d\R \\
\label{e6}
{W'}_\infty^{\rm PC}[\rho] & = & \int \left[C\rho^{3/2}(\R)+D\frac{|\nabla\rho(\R)|^2}{\rho^{7/6}(\R)}\right]d\R\ , 
\end{eqnarray}
with $A=-9(4\pi/3)^{1/3}/10$, $B=3[3/(4\pi)]^{1/3}/350$, $C=\sqrt{3\pi}/2$, and
$D=-0.028957$. 
For small atoms, it has been shown that these PC approximations provide energies quite close to the exact SCE values \cite{seidl07,gorigiorgi09}. More recently, new approximate functionals inspired to the SCE mathematical structure have been also proposed and tested \cite{WagGor-PRA-14,BahZhoErn-JCP-16,VucGor-JPCL-17}. They retain the non-locality of SCE by using as key ingredient some integrals of the density, and their implementation in a self-consistent scheme is still the object of on-going work. 

Different interpolation formulas have been employed to obtain several
XC functionals based on the adiabatic connection formalism 
\cite{isi,isierr,seidl00,gorigiorgi09,SeiPerLev-PRA-99,liu09,ernzrehof99}.
Some of these functionals have been recently tested
against realistic
physical chemistry problems in order to investigate their performance and understand
the corresponding limitations \cite{fabiano16,isigold,vuckovic18}. These functionals are all size-extensive but not size consistent when a system dissociates into fragments of different species. However, it has been recently shown that this size consistency error can be easily corrected at no additional computational cost \cite{vuckovic18}.
All these tests have concerned only the quality of the computed energies, 
whereas no information has been gathered on the XC potentials 
delivered by the functionals.

Actually, the XC potential is a very important quantity since it
enters directly in the Kohn-Sham equations determining the quality
of the Kohn-Sham orbitals and energies as well as the features of
the electron density \cite{Jankowski-2005,ls2,grabowski11,grabowski14, BuiBaeSni-PRA-89,  UmrGon-PRA-94, GriLeeBae-JCP-94, FilGonUmr-INC-96, BaeGri-PRA-96, GriLeeBae-JCP-96, TemMarMai-JCTC-09, HelTokRub-JCP-09,RyaKohSta-PRL-15,OspRyaSta-JCP-17, KohPolSta-PCCP-16, HodRamGod-PRB-16, YinBroLopVarGorLor-PRB-16, BenPro-PRA-16, RyaOspSta-JCP-17}. 
Thus, any accurate XC approximation should be
able to yield not only precise energies at given densities but also
accurate XC potentials. However, this fact is generally overlooked in most
investigations of XC approximations, which mainly focus on the energy
properties only.
On the other hand, a few studies 
\cite{grabowski11,grabowski13,grabowski14_2,smiga16,medvedev49,brorsen17,korth17,mezei17,BuiBaeSni-PRA-89,  UmrGon-PRA-94, GriLeeBae-JCP-94, FilGonUmr-INC-96, BaeGri-PRA-96, GriLeeBae-JCP-96}
have considered the problem of the
XC potential showing that an accurate description of both the
energy and the potential can be usually achieved only by high-rung
approximations (although some important exceptions can be found at the
meta-GGA level of theory \cite{dellasala15,constantin16}).
The ability of an XC functional to describe correctly the XC potential
is therefore a significant problem for DFT development.

In this paper, we consider some relevant XC functionals obtained by interpolating between the two limits of Eqs.~(\ref{eq:lambda0})-(\ref{eq:lambdainf}), studying their ability to
describe the XC potential. Obviously, this potential also depends on how the $\lambda\to\infty$ limit of Eq.~(\ref{eq:lambdainf}) is approximated. We thus start by comparing the functional derivative of the PC model for the leading term,
$W_\infty^{\rm PC}[\rho]$ of Eq.~(\ref{e6}), with the one from the exact SCE formalism for small atoms. Since the PC model is a GEA (and not a GGA), it diverges far from the nucleus. However, we find that in the energetically important region, the PC functional derivative is a rather good approximation of the SCE potential, at least when evaluated on a given reference density (self-consistently the two will give very different results \cite{MalGor-PRL-12,MalMirCreReiGor-PRB-13,MenMalGor-PRB-14}).
Moreover, because (due to the $\lambda\to 0$ expansion) all the considered approximations are
complicated non-linear functionals of the Kohn-Sham orbitals and eigenvalues, full
self-consistent calculations have not been possible, regardless of how the $\lambda\to\infty$ limit is treated. Thus, as explained in 
the next section, the XC potentials have been computed for fixed reference 
densities and compared to accurate reference potentials as well as to
the second-order G\"orling-Levy one. This approach was already successfully utilized
in some studies \cite{One-stepIG,ls2} to investigate the XC potentials properties. Although the procedure does not provide access to the final self-consistent Kohn-Sham orbitals and density, 
it allows anyway to study the quality of the potential and the ability
of each functional to reproduce its most relevant features.
This work is therefore a fundamental first step towards the possible full self-consistent
implementation of adiabatic-connection-based XC functionals.

\section{Potentials from Adiabatic Connection Models}
In this work we consider XC functionals based on different adiabatic 
connection models (ACMs) that use 
the input quantities of Eqs.~(\ref{eq:lambda0})-(\ref{eq:lambdainf}), namely
the exact exchange $E_x$, the second-order G\"orling-Levy correlation energy
$E_c^{\rm GL2}$, the strong-interaction limit of the density-fixed 
adiabatic connection integrand $W_\infty$, and possibly the zero-point term 
$W'_\infty$.
In a compact notation, the XC functionals can be denoted as
\begin{equation}\label{e7}
E_{xc}^{\rm ACM} = f^{\rm ACM}(E_x,E_c^{\rm GL2},W_\infty,W'_\infty)\ ,
\end{equation}
where $f^{\rm ACM}$ is an appropriate non-linear function for the
given ACM model. The exact expressions of the $f^{\rm ACM}$ functions
corresponding to the functionals considered in this work,
namely ISI \cite{isi}, revISI \cite{gorigiorgi09}, 
SPL \cite{SeiPerLev-PRA-99}, and LB \cite{liu09}, are given in Appendix.
Notice that ISI and revISI depend on all four input quantities
while SPL and LB do not depend on $W'_\infty$.
The XC potential corresponding to the
functionals defined by Eq.~(\ref{e7}) is
\begin{eqnarray}
\label{eq8}
v_{xc}^{\rm ACM}(\R) & \equiv & \funcder{E_{xc}^{\rm ACM}}{\rho(\R)} = \\
\nonumber
& = & D^{\rm ACM}_{E_x}\funcder{E_x}{\rho(\R)} + D^{\rm ACM}_{E_c^{\rm GL2}}\funcder{E_c^{\rm GL2}}{\rho(\R)} +\\
\nonumber
&& + D^{\rm ACM}_{W_\infty}\funcder{W_\infty}{\rho(\R)} + D^{\rm ACM}_{W'_\infty}\funcder{W'_\infty}{\rho(\R)} \\
\nonumber
& = & D^{\rm ACM}_{E_x}v_x(\R) + D^{\rm ACM}_{E_c^{\rm GL2}}v_c^{\rm GL2}(\R) +\\
\nonumber
&& + D^{\rm ACM}_{W_\infty}\funcder{W_\infty}{\rho(\R)} + D^{\rm ACM}_{W'_\infty}\funcder{W'_\infty}{\rho(\R)}\ ,
\end{eqnarray}
where
\begin{eqnarray}
\label{e9}
&&D^{\rm ACM}_{E_x} = \deri{f^{\rm ACM}}{E_x}\; , \; D^{\rm ACM}_{E_c^{\rm GL2}} = \deri{f^{\rm ACM}}{E_c^{\rm GL2}}\; , \\
\label{e10}
&& D^{\rm ACM}_{W_\infty} = \deri{f^{\rm ACM}}{W_\infty}\; , \; D^{\rm ACM}_{W'_\infty} = \deri{f^{\rm ACM}}{W'_\infty}\; ,
\end{eqnarray}
and we used the short-hand notations
\begin{equation}
v_x(\R) = \funcder{E_x}{\rho(\R)} \; , \; v_c^{\rm GL2}(\R)=\funcder{E_c^{\rm GL2}}{\rho(\R)}\ .
\end{equation}
For the functionals not depending on $W'_\infty$, i.e. SPL and LB,
we have $D^{\rm ACM}_{W'_\infty} =0$. The derivatives of Eqs.~(\ref{e9}) and (\ref{e10}) are straightforward
once the function $f^{\rm ACM}$ is fixed. The potentials $\delta W_\infty/\delta\rho$ and $\delta W'_\infty/\delta\rho$
depend on how the $\lambda\to\infty$ limit is treated. For the PC model, they are standard gradient expansion functional derivatives (see Appendix). For the exact case, they can be computed by integrating the SCE force equation \cite{seidl07,MalGor-PRL-12,MalMirCreReiGor-PRB-13}, see the subsection on the potentials for the strong-interaction limit below.

\subsection{Potentials for the weak-interaction limit}
The calculations of $v_x$ and $v_c^{\rm GL2}$ require some
attention. In this work we have
used the optimized effective potential (OEP) method \cite{grabowski14_2,gorling05}. 
Then, the potential
$v_x$ or $v_c^{\rm GL2}$ (denoted with general notation $v$ in Eq. (\ref{e13})) 
is given by the solution of the integral equation
\begin{equation}\label{e13}
\int \chi(\R,\R')v(\R')d\R' = \Lambda(\R)\ ,
\end{equation}
where $\chi(\R,\R')$ is the static linear-response function and
\begin{small}
\begin{eqnarray}
\nonumber
\Lambda(\R) & = & \sum_p\left[\int\phi_p(\R)\funcder{E}{\phi_p(\R)}\sum_{q\neq p}\frac{\phi_q(\R)\phi_q(\R')}{\epsilon_p-\epsilon_q}d\R' \right.\\
&&\quad\quad + \left.\funcder{E}{\epsilon_p}|\phi_p(\R)|^2\right]\ ,
\end{eqnarray}
\end{small}
with $E$ being either the exact exchange or the second-order G\"orling-Levy
correlation energies, $\phi_p$ and $\epsilon_p$ being the Kohn-Sham orbitals
and orbital energies, respectively. A similar approach was used in
Ref.~\cite{DHOEPSmiga2016}, leading to the fully self-consistent
solution of double-hybrid functionals.

\subsection{Computational details}
In our study we have considered several atoms, He, Be, Ne, and Ar,
the H$_2^-$ and F$^-$ ions, as well as the H$_2$ and N$_2$ molecules.
For all these systems we have determined the self-consistent
exact exchange orbitals and density.
Exact exchange potential calculations have been carried out
with a locally modified version of the ACESII code \cite{acesII}
using an uncontracted cc-pVTZ
basis set \cite{dunning89} for H$_2$ and N$_2$, 
a 20s10p2d basis set \cite{grabowski02} for He, 
an uncontracted ROOS-ATZP basis set \cite{widmark1990} for H, Be, Ne, and F, 
and for the Ar atom a modified basis set combining $s$- and
$p$-type basis functions from the uncontracted ROOS-ATZP \cite{widmark1990} 
with $d$- and $f$-type functions coming from the uncontracted
aug-cc-pwCVQZ basis set \cite{grabowski13,peterson02}.
The choice of the basis set
was mostly dictated by the need to ensure the best possible
expansion for both the wave functions and the OEP potential.
For more details see Refs. \cite{smiga16,grabowski14_2}.
SUccessively we have computed, in a post-SCF fashion, 
the XC potentials of the various ACM XC functionals
(ISI \cite{isi}, revISI \cite{gorigiorgi09}, 
SPL \cite{SeiPerLev-PRA-99}, and LB \cite{liu09}; see Appendix)
using Eq. (\ref{eq8}). Similarly we have computed, in a one-step procedure
based on exact exchange orbitals, the GL2 potential as well as the
other OEP correlation potentials (OEP2-sc,OEP2-SOSb).
In all these potentials the single excited term has been neglected; anyway this 
is expected to yield a negligible effect on the final result 
\cite{grabowski14_2}.

\subsection{Potentials for the strong-interaction limit}
The exact (or very accurate \cite{SeiDiMGerNenGieGor-arxiv-17}) functional derivative of $W_\infty[\rho]$ can be 
computed for spherically symmetric densities using the formalism and the procedure described in 
Refs.~\cite{seidl07,MenMalGor-PRB-14,SeiDiMGerNenGieGor-arxiv-17}. 
The potential is obtained by integrating numerically the force equation 
\begin{equation}
\nabla v_{\rm SCE}[\rho](\R)=-\sum_{i= 2}^N \frac{\R-\fv_i(\R;[\rho])}{|\R-\fv_i(\R;[\rho])|^3},
\label{eq:vSCE}
\end{equation}
where the co-motion functions $\fv_i(\R;[\rho])$, which are highly non-local density functionals, 
portray the strictly-correlated regime, determining the positions of $N-1$ electrons as functions of 
the position $\R$ of one of them \cite{seidl07,MenMalGor-PRB-14,SeiDiMGerNenGieGor-arxiv-17}, and the boundary condition $v_{\rm SCE}[\rho](|\R|\to\infty)=0$ is used.
Once $v_{\rm SCE}$ of Eq.~(\ref{eq:vSCE}) has been computed, we then have the exact relation 
\cite{MalGor-PRL-12,SeiDiMGerNenGieGor-arxiv-17}
\begin{equation}
\frac{\delta W_\infty[\rho]}{\delta\rho(\R)}\equiv v_{xc}^\infty(\R)=v_{\rm SCE}(\R)-v_{\rm H}(\R), 
\label{eq:vxcSCE}
\end{equation}
where $v_{\rm H}(\R)$ is the Hartree potential. 

We have computed $v_{xc}^{\infty}(\R)$ of Eqs.~(\ref{eq:vSCE})-(\ref{eq:vxcSCE}) for H$^{-}$, Be, and Ne, 
for the same densities as described in the Computational Details Section, using the co-motion functions described in 
Refs.~\cite{seidl07,MenMalGor-PRB-14,SeiDiMGerNenGieGor-arxiv-17}, 
which are exact for $N=2$ and very accurate (or exact) for $N>2$ \cite{SeiDiMGerNenGieGor-arxiv-17}. 
Moreover, even when these co-motion functions are not optimal, the functional derivative 
of the corresponding $W_\infty[\rho]$ still obeys Eqs.~(\ref{eq:vSCE})-(\ref{eq:vxcSCE}) \cite{SeiDiMGerNenGieGor-arxiv-17}. 

\begin{figure}[t]
\includegraphics[width=0.9\columnwidth]{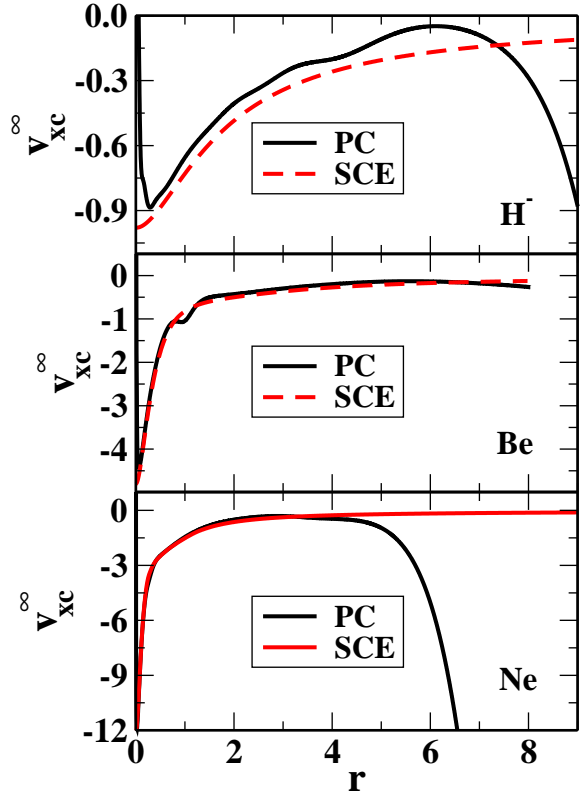}
\caption{\label{fig:vxcSCE} Comparison between the exact (or very accurate) functional derivative of the leafing term $W_\infty[\rho]$ in Eq.~(\ref{eq:lambdainf}), computed with the strictly-correlated electrons (SCE) formalism of Eqs.~(\ref{eq:vSCE})-(\ref{eq:vxcSCE}) \cite{seidl07,MenMalGor-PRB-14,SeiDiMGerNenGieGor-arxiv-17}, and the functional derivative of the GEA approximation obtained from the PC model of Eq.~(\ref{e5}).}
\end{figure}

In Figure~\ref{fig:vxcSCE}, we compare these exact (or very accurate) $v_{xc}^{\infty}(\R)$ 
with the functional derivative of $W_\infty^{\rm PC}[\rho]$ for the three species. 
We see that, as anticipated, since the PC model is a GEA, it diverges at large internuclear distances, 
tending to minus infinity (see Appendix for more details), a feature that would further prevent to perform self-consistent calculations. 
The exact $v_{xc}^{\infty}(\R)$, instead, displays the correct asymptotic behavior 
$\propto -1/r$ \cite{seidl07,MalGor-PRL-12}. Nonetheless, we see that in the region where the density 
is significantly different from zero, the PC model provides a very decent approximation to 
the exact  $v_{xc}^{\infty}(\R)$. In the next sections, we will then use the much cheaper PC potentials, 
since we will always compute them on a reference density, where they seem to give rather reasonable 
results (with the exception of the region close and far from the nucleus). This choice has been further 
validated by comparing for a few cases the potentials obtained from the SPL and LB models using the 
PC and SCE functional derivatives: in all cases the differences between the two have been found to be very small, 
of course with the exception of the asymptotic region far from the nucleus.

\begin{figure}[t]
\includegraphics[width=\columnwidth]{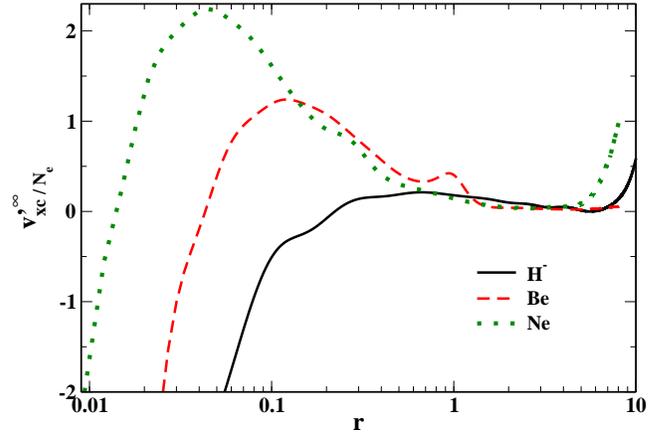}
\caption{\label{fig_w1} Functional derivative of ${W'}_\infty[\rho]$, divided by the number of electrons, obtained using the GEA approximation based on the PC model of Eq.~(\ref{e6}), for several atoms.}
\end{figure}

There is at present no exact result available for the functional derivative of the zero-point term $W_\infty'[\rho]$, 
which is the object of an on-going investigation. For this reason, we will use again the PC model, 
which also yields a potential that diverges, this time going to plus infinity, but only rather 
far from the nucleus (see Fig. \ref{fig_w1} and Appendix for more details).

\section{Results}
\begin{figure}[t]
\includegraphics[width=0.95\columnwidth]{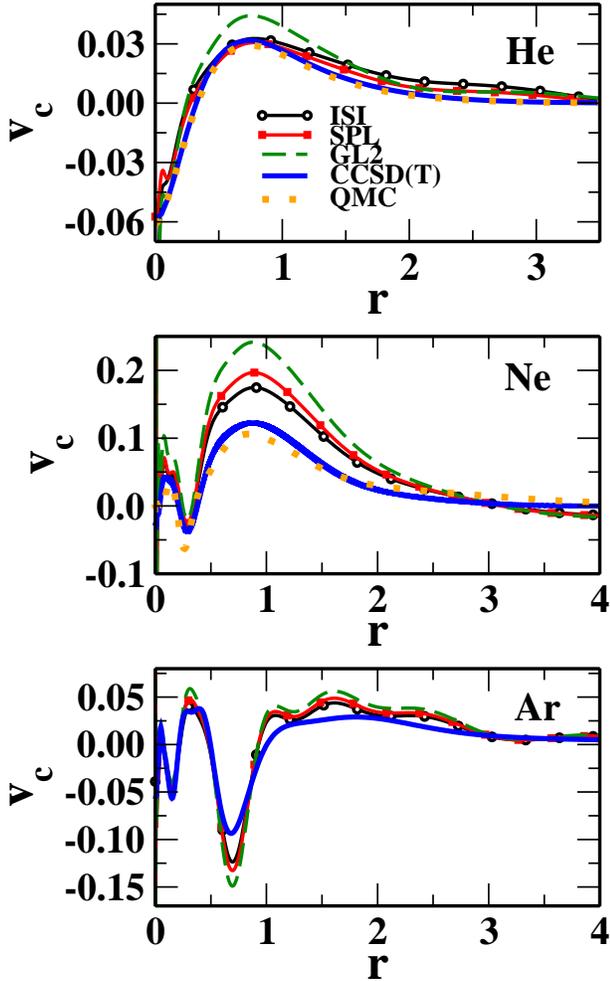}
\caption{\label{fig_pot} Correlation potentials obtained with different methods
for several atoms.}
\end{figure}

In Figure \ref{fig_pot} we show the correlation potentials
corresponding to different XC functionals (on the scale of the
plot ISI and revISI are hardly distinguishable, therefore only the ISI
curve has been shown; the same applies to SPL and LB). 
We recall that all the
methods considered in this work include the exchange contribution
exactly; therefore only the correlation part of the potential 
($v_c^{\rm ACM} = v_{xc}^{\rm ACM}-v_x$) is
of interest for an assessment of the methods.
For comparison we have also reported the correlation potential
obtained via direct inversion of the coupled cluster single double with
perturbative triple (CCSD(T)) density\cite{wu:2003:wy} as well as
the one obtained from quantum Monte Carlo (QMC) calculations
\cite{umrigar:1994:EXACT,Filippi:1996:EXACT}. Both potentials are very accurate and are assumed
here as benchmark references. Note anyway that these potentials
do not stem from the exact exchange densities used to generate the
ACMs potentials, but correspond to self-consistent CCSD(T) or QMC densities.
Nevertheless, we expect the difference due to this issue to be almost
negligible for the purpose of this work.

\begin{figure}[t]
\includegraphics[width=0.95\columnwidth]{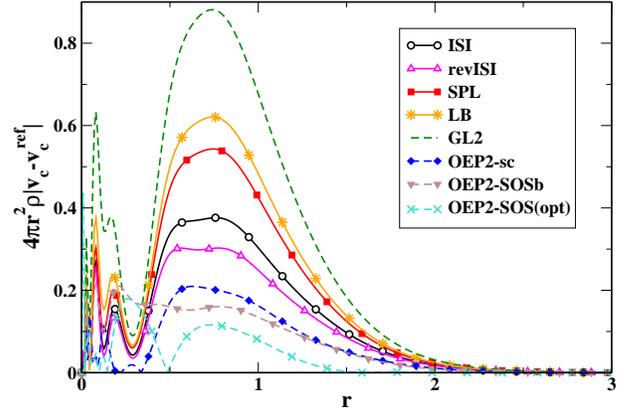}
\caption{\label{fig_neerrors}Radial density-weighted absolute errors (see the integrand of Eq. (\ref{e14})) for various potentials based on adiabatic connection models as well as for GL2 and several second-order OEP potentials, in the case of Ne atom.}
\end{figure}

The plots of Fig. \ref{fig_pot} show that, in all cases, the 
various potentials have quite similar shapes. 
This indicates that for all the ACMs the correlation potential is
physically meaningful and describes the main features of the
exact potential. Moreover, the
ACMs potentials are generally closer to the reference ones
than GL2. This latter situation can be measured quantitatively
considering the integral error
\begin{equation}\label{e14}
\Delta v_c^{\rm ACM} = \int\rho(\R)|v_c^{\rm ACM}(\R)-v_c^{Ref}(\R)|d\R\ ,
\end{equation}
where we use the CCSD(T) potential as reference ($Ref$)
and $\rho$ is the exact exchange self-consistent density
(i.e. the one used to compute $v_c^{\rm ACM}$).
The corresponding values are reported in Table \ref{tab1},
whereas a plot of the integrands for the Ne atom case
is shown in Fig. \ref{fig_neerrors}.
\begin{table}
\caption{\label{tab1}Integral error ($\Delta v_c^{\rm ACM}$; see Eq. (\ref{e14})) for different potentials.}
%\begin{ruledtabular}
\begin{tabular}{lccc}
\hline\hline
 & He & Ne & Ar \\
\hline
ISI    & 0.0088 & 0.3734 & 0.1718 \\  
revISI & 0.0113 & 0.3136 & 0.1583 \\
SPL    & 0.0060 & 0.5083 & 0.2123 \\
LB     & 0.0070 & 0.5778 & 0.2258 \\
GL2    & 0.0212 & 0.8151 & 0.2886 \\
OEP2-sc& 0.0133 & 0.1900 & 0.1062 \\
OEP2-SOSb& 0.0170 & 0.2140 & 0.1710 \\
OEP2-SOS(opt)& 0.0102 & 0.1127 & 0.2339 \\
\hline\hline
\end{tabular}
%\end{ruledtabular}
\end{table}
These data confirm that the ACMs provide a significant improvement
over GL2, yielding values of $\Delta v_c^{\rm ACM}$ that are in most cases almost one half
that those of GL2. Moreover, the table reports also the integral errors for some
accurate second-order OEP methods, namely OEP2-sc \cite{bartlett:2005:abinit2,IGccpt2},
OEP2-SOSb \cite{grabowski14_2,smiga16} and OEP2-SOS(opt) \cite{smiga16},
indicating that the ACMs potentials are competitive with the best available
OEP methods. Finally,  we find that in general ISI and revISI are
slightly better than SPL and LB, most probably due to inclusion of the
zero-point term ${W'}_{\infty}[\rho]$.

An additional indication of the quality of the ACMs potentials can come by the
analysis of the effect they have on the Kohn-Sham orbital energies.
In particular, we have considered the first-order perturbative variation of the energy of
the highest occupied molecular orbital (HOMO), i.e. 
$\langle H|v_c^\mathrm{AMC}|H\rangle$ with $|H\rangle$ being the HOMO state, due to the
application of the potential $v_c^\mathrm{AMC}$. These data are reported in
Table \ref{homo_tab} together with some reference values.
\begin{table*}
\caption{\label{homo_tab} First order variation of the energy (eV) of the Kohn-Sham
  highest occupied molecular orbital (HOMO) as due to the application of different
  correlation potentials. The right part of the table reports, for
  comparison, the difference between the self-consistent HOMO energy computed with some
  correlated methods and the one obtained at the exact exchange level ($\epsilon_H-\epsilon_H^\mathrm{OEPx}$). Reference values are obtained from inverted CCSD(T) calculations.}
%\begin{ruledtabular}
\begin{tabular}{lcccccccccr}
\hline\hline
 & \multicolumn{5}{c}{$\langle H|v_c^\mathrm{AMC}|H\rangle$} &$\;\;$ & \multicolumn{4}{c}{$\epsilon_H-\epsilon_H^\mathrm{OEPx}$}  \\
\cline{2-6}\cline{8-10}
 & ISI & revISI & SPL & LB & GL2 & & GL2 & OEP2-sc & OEP2-SOSb & Ref. \\
\hline
He & 0.576 & 0.610 & 0.505 & 0.551 & 0.746 & & 0.751 & 0.427 & 0.547 & 0.419 \\
Ne & 3.211 & 3.026 & 3.627 & 3.826 & 4.528 & & 5.491 & 3.012 & 2.416 & 1.935 \\
Ar & 0.573 & 0.538 & 0.652 & 0.680 & 0.767 & & 1.137 & 0.740 & 0.558 & 0.672 \\
\hline\hline
\end{tabular}
%\end{ruledtabular}
\end{table*}
These results confirm the trends observed for the $\Delta v_c^{\rm ACM}$, showing that
all the ACMs potentials yield HOMO energy variations in line with OEP2-sc and
OEP2-SOSb and close to the reference values; on the other hand GL2 yields quite overestimated
HOMO energy variations.

\begin{figure}[t]
\includegraphics[width=0.95\columnwidth]{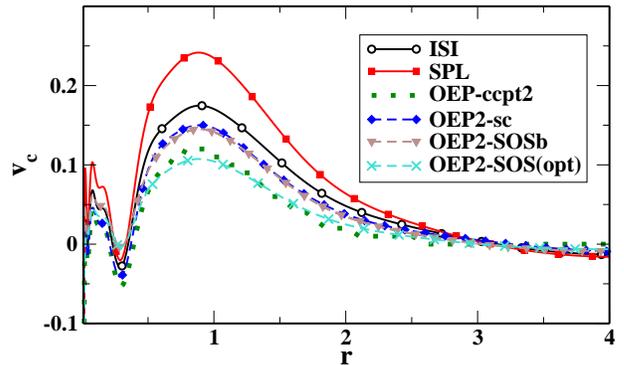}
\caption{\label{fig_oep_atom}Correlation potentials obtained with different correlated OEP methods and two ACMs (ISI, SPL)
for the neon atom. }
\end{figure}

We remark that the improvement found for the ACMs with respect to
GL2 is particularly significant because it mainly corresponds to a reduction
of the GL2 overestimation of the potential in the outer valence region.
This feature is in fact one the main limitations that GL2 experiences in
OEP Kohn-Sham calculations, which leads to several problems
including, sometimes, the impossibility to converge OEP-GL2 Kohn-Sham 
self-consistent calculations. Indeed, several
modifications of OEP-GL2, such as OEP2-sc,
OEP2-SOSb, and OEP2-SOS(opt), have been developed to account for this problem 
\cite{bartlett:2005:abinit2,grabowski13,grabowski14_2,smiga16,CA-GL2}.
Thus, the partial correction of this drawback from the
ACMs is a very promising feature of these functionals
which allows them to yield potentials close to the accurate 
OEP ones
(see Fig. \ref{fig_oep_atom} where we compare, for the Ne atom, ISI and SPL with OEP2-sc,
OEP2-SOSb, OEP2-SOS(opt) as well as with OEP-ccpt2 \cite{IGccpt2}).

\begin{figure}[t]
\includegraphics[width=0.95\columnwidth]{be_oepx.eps}
\includegraphics[width=0.95\columnwidth]{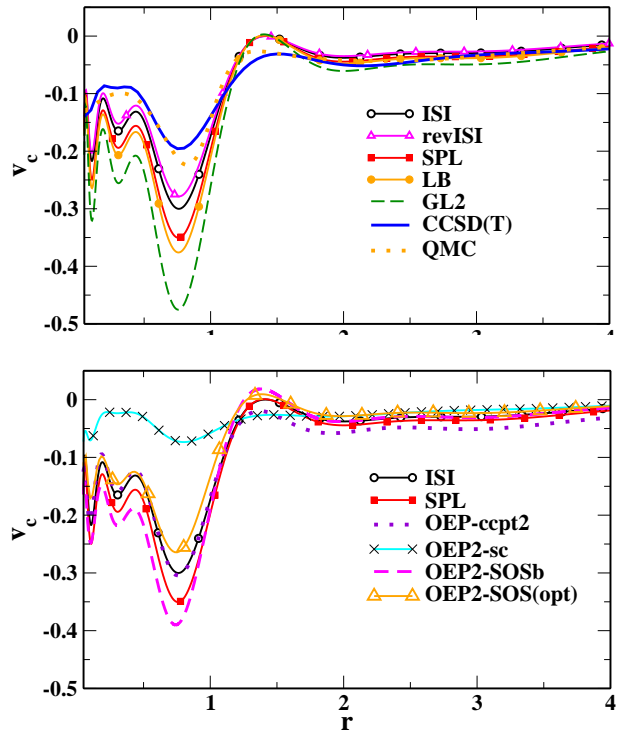}
\caption{\label{fig_be}Correlation potentials obtained with different methods
for the beryllium atom.}
\end{figure}

As a further example we consider the case of the beryllium atom
which is known to be a rather extreme case where the GL2 potential 
performs poorly, overestimating the correct correlation potential,
such that self-consistent OEP-GL2 calculations fail to converge.
Thus, in Fig. \ref{fig_be} we report the various correlation potentials
computed for beryllium (note that in the bottom panel we report also 
the accurate second-order OEP potentials for comparison). 
Inspection of the plots shows that also in this difficult case
the ACMs potentials, especially ISI and revISI, improve
substantially over GL2 being comparable to the OEP2-SOSb and OEP2-SOS(opt)
approaches. Note also that the accurate OEP2-sc method instead yields an
underestimation of the correlation potential of the beryllium atom.

The analysis discussed above, evidenced the similarity
of the ACMs potentials with scaled GL2 ones \cite{grabowski13,grabowski14_2,smiga16}. 
This finding is not completely surprising if we inspect the magnitudes of the 
different contributions forming the ACM potential in Eq. (\ref{eq8}), i.e. $D^{\rm ACM}_{E_x}$, 
$D^{\rm ACM}_{E_c^{\rm GL2}}$, $D^{\rm ACM}_{W_\infty}$, and $D^{\rm ACM}_{W'_\infty}$ 
(see Table  \ref{tab_vders}). 
\begin{table}
\caption{\label{tab_vders} Values of the partial derivatives appearing in Eq. (\ref{eq8}) for the various ACM correlation functionals of different atoms.}
%\begin{ruledtabular}
\begin{tabular}{llcccc}
\hline\hline
 & & $D^{\rm ACM}_{E_x}-1$ & $D^{\rm ACM}_{E_c^{\rm GL2}}$ & $D^{\rm ACM}_{W_\infty}$ & $D^{\rm ACM}_{W'_\infty}$ \\
\hline
He & ISI    & -0.029 & 0.642 & 0.029 & 0.006 \\
   & revISI & -0.036 & 0.615 & 0.036 & 0.009 \\
   & SPL    & -0.014 & 0.699 & 0.014 & - \\
   & LB     & -0.012 & 0.757 & 0.012 & - \\
   &        &        &       &       & \\
Be & ISI    & -0.030 & 0.617 & 0.030 & 0.005 \\
   & revISI & -0.038 & 0.570 & 0.038 & 0.007 \\
   & SPL    & -0.011 & 0.726 & 0.011 & - \\
   & LB     & -0.009 & 0.782 & 0.009 & - \\
   &        &        &       &       & \\
Ne & ISI    & -0.014 & 0.724 & 0.014 & 0.002 \\
   & revISI & -0.018 & 0.683 & 0.018 & 0.003 \\
   & SPL    & -0.005 & 0.819 & 0.005 & - \\
   & LB     & -0.004 & 0.869 & 0.004 & - \\
   &        &        &       &       & \\
Ar & ISI    & -0.007 & 0.079 & 0.007 & 0.001 \\
   & revISI & -0.010 & 0.751 & 0.010 & 0.001 \\
   & SPL    & -0.002 & 0.873 & 0.002 & - \\
   & LB     & -0.002 & 0.904 & 0.002 & - \\
\hline\hline
\end{tabular}
%\end{ruledtabular}
\end{table}
The data show in fact that the main contribution to the
potential of any of the ACMs is a scaled GL2 component,
with a magnitude of about 70\%. This finding traces back to
the fact that the ACMs are effectively all-order renormalizations
of the density functional perturbation theory \cite{isi},
thus they basically perform a ``rescaling'' of the GL2
correlation contribution. This feature puts these potentials not only in
relation with the spin-opposite-scaled OEP methods (OEP2-SOS), as mentioned above, but also with the
recently developed self-consistent OEP double-hybrid \cite{DHOEPSmiga2016} method, 
which also shows a similar signature.
Then, the success of the later methods in self-consistent calculations can be considered as
promising indicator for the quality of the ACM potentials and the possible success of future self-consistent
calculations based on the ACM functionals.

\begin{figure}[t]
\includegraphics[width=0.95\columnwidth]{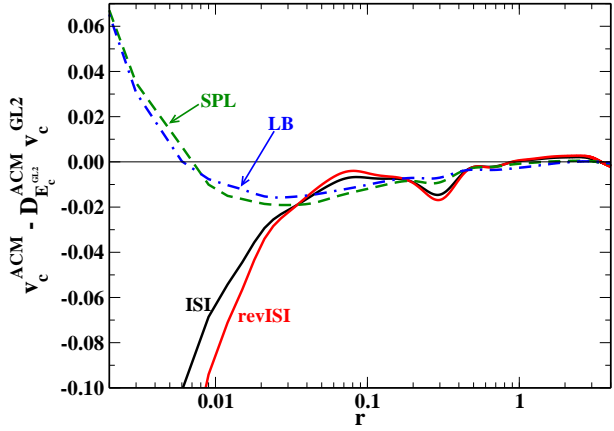}
\caption{\label{fig_diff}Difference between the correlation potentials of various adiabatic connection models and the scaled second-order G\"orling Levy correlation potential ($v_c^{\rm ACM}-D^{\rm ACM}_{E_c^{\rm GL2}}v_c^{\rm GL2}$) for the neon atom.}
\end{figure}

We remark anyway that, despite the scaled GL2 is the main component
of the correlation potential of ACMs, other contributions
may also be present with a non negligible effect.
This can be seen, for example, in Fig. \ref{fig_diff} where we plot
the quantity $v_c^{\rm ACM}-D^{\rm ACM}_{E_c^{\rm GL2}}v_c^{\rm GL2}$ for the Ne atom.
The plot shows that all the ACM potentials
show relevant features in the core regions (i.e. 
$v_c^{\rm ACM}-D^{\rm ACM}_{E_c^{\rm GL2}}v_c^{\rm GL2}\ne 0$), 
that possibly originate 
from the larger correlation effects felt by core electrons,
implying a larger contribution from the $D^{\rm ACM}_{W_\infty}$
component (see also the subsection Potentials for the strong-interaction limit).
This is more evident for ISI and revISI, whereas,
the SPL and LB correlation potentials
are slightly closer to scaled GL2 potentials.

\begin{figure}[t]
\includegraphics[width=0.95\columnwidth]{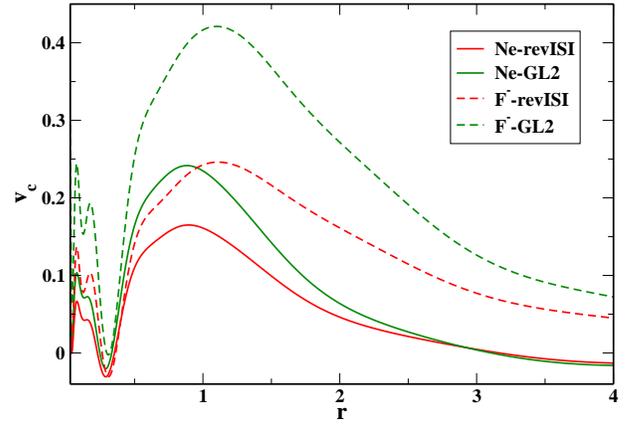}
\caption{\label{fig_ion}GL2 and revISI correlation potentials for the Ne atom and the F$^-$ anion.}
\end{figure}

It is worth to note also that the ``renormalization'' effect
of the  ACM correlation potentials will increase when
systems with stronger correlation are considered. 
Unfortunately, for such systems the OEP equation needed to generate
the GL2 potential cannot be generally solved; thus we had
to limit our investigation only to the simple case of the F$^-$
anion, to be compared with the Ne atom. This comparison
is shown in Fig. \ref{fig_ion} for the case of the revISI
correlation potential (other ACMs behave similarly).
From the plot it can be seen that for the F$^-$ anion, 
where correlation effects are slightly larger than in Ne, 
indeed a greater difference between the revISI and the
GL2 potentials is found. This fact is also confirmed by the
computed value of $D^{\rm ACM}_{E_c^{\rm GL2}}$ that for F$^-$ is
0.583 (to be compared with 0.683 in Ne, see Table \ref{tab_vders});
note also that $D^{\rm ACM}_{W_\infty}$ is 0.030 for F$^-$ while
it is 0.018 for the neon atom.

\begin{figure}[t]
\includegraphics[width=0.95\columnwidth]{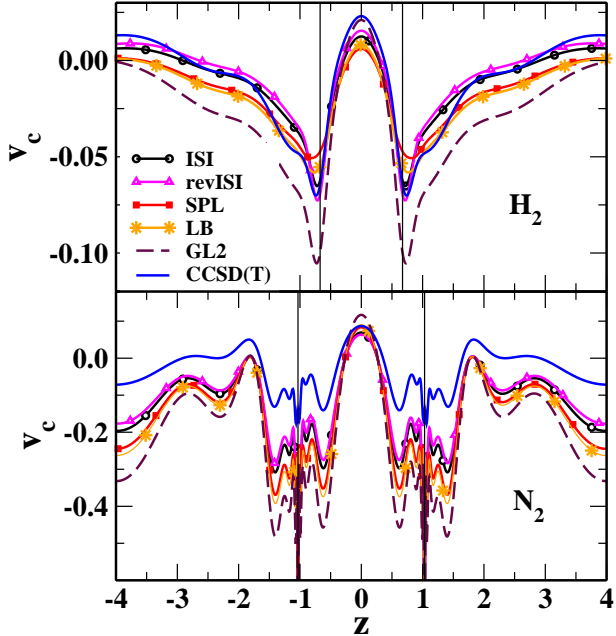}
\caption{\label{fig_mol}Correlation potentials of different adiabatic connection models as well of the second order G\"orling-Levy (GL2) correlation for the H$_2$ (top) and N$_2$ (bottom) molecules, plotted along the bond axis. The vertical lines indicate the positions of the atoms in each molecule. }
\end{figure}

To conclude, in Fig. \ref{fig_mol} we report the correlation potentials
computed for two simple molecules, namely H$_2$ and N$_2$. 
In these cases we have also used as a reference the correlated potential
obtained from relaxed CCSD(T) density observing the same qualitative behavior as already found in atoms.
Thus, all the ACMs provide a reduction of the correlation potential with
respect to GL2. In particular, ISI and revISI provide a slight larger
reduction than SPL and LB.
\begin{figure}[t]
\includegraphics[width=0.95\columnwidth]{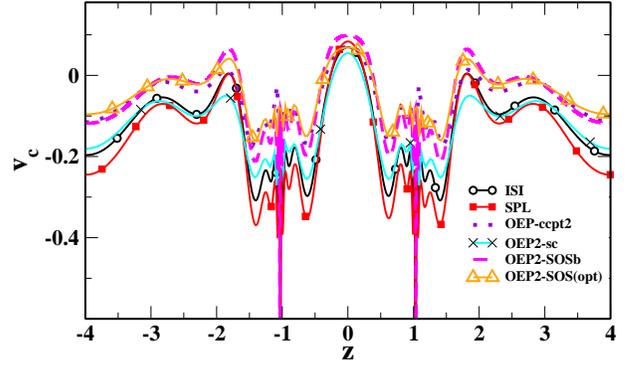}
\caption{\label{fig_oep_molecule}Correlation potentials for nitrogen dimer obtained with  several correlated OEP methods and two ACMs (ISI, SPL).}
\end{figure}
Also here the direct comparison with several correlated OEP potentials made for nitrogen dimer reveals remarkably good performance for all ACMs. The ISI and SPL give almost the same potentials as OEP2-sc which is considered as the state-of-the-art correlated OEP method.

\section{Conclusions}
We have studied the correlation potentials produced by different adiabatic connection
models to investigate whether these methods are able to produce, besides reasonable energies
\cite{fabiano16,isigold,vuckovic18}, also physically meaningful potentials. 
This is, in fact, a fundamental issue in view of a possible
future use of the ACMs in self-consistent calculations.

Our results showed that indeed all the investigated functionals are able to provide 
rather accurate correlation potentials, with ISI and revISI being slightly superior to
SPL and LB. In particular, all the considered correlation potentials display the correct features
of the exact correlation potential and reduce the overestimation behavior which is typical
of the GL2 method \cite{bartlett:2005:abinit2,schweigert06}.
Thus, all in all, the ACM-based correlation potentials are comparable to those produced
by some state-of-the-art optimized effective potential methods 
(OEP2-sc \cite{bartlett:2005:abinit2,IGccpt2},
OEP2-SOS \cite{grabowski14_2,smiga16}).

These results suggest that it might be worth to pursue the realization
of self-consistent calculations using the ACM XC functionals, which
would provide a final assessment of their quality.
However, to reach this goal a few  issues need first to be solved.
In particular, one needs to develop proper density functionals for the $\lambda\to\infty$ limit, in order to replace the gradient expansions of Eqs. (\ref{e5}) and 
(\ref{e6}), which lead to divergences in the potential. The functionals of Refs.~\cite{WagGor-PRA-14,BahZhoErn-JCP-16,VucGor-JPCL-17} are already very good candidate, although they pose new technical problems at the implementation level due to their non-local density dependence. If one wants to stay within gradient approximations, then the PC GEA should be renormalized into GGA's. Moreover,
a proper OEP scheme, similar to the one used for the implementation
of self-consistent double hybrids \cite{DHOEPSmiga2016}, need to be implemented to
deal with the various terms appearing in Eq. (\ref{eq8}), which are both of 
implicit non-local and explicit semilocal types.

\section{Acknowledgements}
S.\'S is grateful to the Polish National Science Center for the partial financial support under Grant No. 2016/21/D/ST4/00903.
P.G.-G. and S.G. acknowledge funding from the European Research Council under H2020/ERC Consolidator Grant corr-DFT (Grant Number 648932), and T.J.D. is grateful to the Vrije Universiteit Amsterdam for a University Research Fellowship.  

\appendix

\section{Adiabatic connection models}
In this work we consider several ACMs. They main features of each one
are described below.

\textbf{ISI functional} \cite{isi}.
The XC energy is
\begin{equation}
E_{xc}^{\rm ISI} = W_\infty + \frac{2X}{Y}\left[\sqrt{1+Y}-1-Z\ln\left(\frac{\sqrt{1+Y}+Z}{1+Z}\right)\right]\ ,
\end{equation}
with
\begin{eqnarray}
&& X = \frac{xy^2}{z^2}\; , \; Y=\frac{x^2y^2}{z^4}\; ,\; Z=\frac{xy^3}{z^3}-1\; ,\\
&& x = -4E_c^{\rm GL2}\; ,\; y=W'_\infty\; , \; z=E_x-W_\infty.
\end{eqnarray}

\textbf{revISI functional} \cite{gorigiorgi09}.
The XC energy is
\begin{equation}
E_{xc}^{\rm revISI} = W_\infty + \frac{b}{\sqrt{1+c}+d}\ ,
\end{equation}
with
\begin{eqnarray}
b & = & -\frac{8E_c^{\rm GL2}(W'_\infty)^2}{(E_x-W_\infty)^2}\ ,\\
c & = & \frac{16(E_c^{\rm GL2}W'_\infty)^2}{(E_x-W_\infty)^4}\ ,\\
d & = & -1 - \frac{8E_c^{\rm GL2}(W'_\infty)^2}{(E_x-W_\infty)^3}\ .
\end{eqnarray}

\textbf{SPL functional} \cite{SeiPerLev-PRA-99}.
The XC energy is
\begin{equation}
E_{xc}^{\rm SPL} = E_x + (E_x-W_\infty)\left[\frac{\sqrt{1+2\chi}-1-\chi}{\chi}\right]\ ,
\end{equation}
with
\begin{equation}
\chi = \frac{2E_c^{\rm GL2}}{W_\infty-E_x}\ .
\end{equation}

\textbf{LB functional} \cite{liu09}.
The XC energy is
\begin{equation}
E_{xc}^{\rm LB} = E_x + \frac{2b}{c}\left[\sqrt{1+c}-\frac{1+c/2}{1+c}-c\right]\ ,
\end{equation}
with
\begin{equation}
b=\frac{E_x-W_\infty}{2}\; , \; c = \frac{8E_c^{\rm GL2}}{5(W_\infty-E_x)}\ .
\end{equation}
%\R

\section{Functional derivative of the PC functionals}
The functional derivative of Eqs. (\ref{e5}) and (\ref{e6}) with respect to the density is
\begin{equation}\label{eq:b1}
\frac{\delta{W}^{\rm PC}_{\infty}[\rho]}{\delta\rho(\R)}=\frac{4A}{3} \rho(\R)^{1/3}-2B\frac{\nabla^{2}\rho(\R)}{\rho(\R)^{4/3}}+\frac{4B}{3}\frac{|\nabla\rho(\R)|^{2}}{\rho(\R)^{7/3}},
\end{equation}
\begin{equation}\label{eq:b2}
\frac{\delta{W}'^{\rm PC}_{\infty}[\rho]}{\delta\rho(\R)}=\frac{3C}{2} \rho(\R)^{1/2}-2D\frac{\nabla^{2}\rho(\R)}{\rho(\R)^{7/6}}+\frac{7D}{6}\frac{|\nabla\rho(\R)|^{2}}{\rho(\R)^{13/6}}.
\end{equation}
In order to study which terms are responsible for the divergence observed (see the subsection Potentials for the strong-interaction limit), we consider the asymptotic expression \cite{KatNav-PNAS-80} for a density of $N$ electrons bound by a total positive charge $Z$, with $r=|\R|$,
\begin{equation}\label{eq:b3}
\rho(r\rightarrow\infty)\sim r^{\beta}e^{-{\alpha}r},
\end{equation}
with $\alpha=2\sqrt{2I}$, $\beta=\frac{Z-N+1}{\sqrt{2I}}-1$, and $I$ the first ionization potential  (e.g. for the Ne atom $\alpha \approx 2.5$ and $\beta \approx - 0.2$).
By plugging Eq.~(\ref{eq:b3}) into Eq.~(\ref{eq:b1}), one sees that the second and third terms go to leading order like $ r ^{- \frac{\beta}{3}} e^{\frac{\alpha}{3} r}$; however the second term is multiplied by a negative constant, namely $- 2 B$, which determines the divergence to $-\infty $ over the third term, whose prefactor is $\frac{4 B}{3}$.
Analogously, for Eq.~(\ref{eq:b2}), one sees that the second and third terms, again, diverge with same leading order, which, in this case, is $r ^{- \frac{\beta}{6}} e^{\frac{\alpha}{6} r}$. In this latter case, however, the second term is multiplied by the positive constant $- 2 D$, which determines the divergence to $+ \infty$ over the third term, whose prefactor is $\frac{7 D}{6}$.
Since all the ACM potentials calculated in this work make use of the PC  approximation for the strong-interaction limit terms, they all show a negative divergence, appearing `earlier' in the SPL and LB than in the ISI and revISI models (as these latter also include the ingredient $\frac{\delta{W}'^{\rm PC}_{\infty}[\rho]}{\delta\rho(\R)}$, which partially compensates the asymptotically dominant term). Moreover, while this analysis applies to the exact asymptotic density behavior, the use of gaussian basis sets worsens the divergence at large $r$.

\bibliography{isipot,bib_clean}

\end{document}